\documentclass[aps,prb,reprint,amsmath,amssymb,superscriptaddress,showpacs]{revtex4-2}
\newcommand{\del}[1]{\textcolor{red}{\sout{#1}}}
\newcommand{\add}[1]{\textcolor{red}{#1}}
\renewcommand{\del}[1]{}
\renewcommand{\add}[1]{#1}
\usepackage{graphicx}
\usepackage{dcolumn}
\usepackage{bm}
\usepackage[colorlinks,linkcolor=blue,anchorcolor=blue,citecolor=blue,urlcolor=blue,filecolor=blue,menucolor=blue,runcolor=blue]{hyperref}
\usepackage[normalem]{ulem}  
\usepackage{xcolor}          

\begin{document}
\title {Infinite Magnetoresistance and Vortex Coupling in the Pb/BSCCO Heterostructure}
\author{Weifan Zhu}
\affiliation{Center for Correlated Matter and School of Physics, Zhejiang University, Hangzhou 310058, China}
\author{Jiamin Yao}
\affiliation{Center for Correlated Matter and School of Physics, Zhejiang University, Hangzhou 310058, China}
\author{Shuntianjiao Ling}
\affiliation{Center for Correlated Matter and School of Physics, Zhejiang University, Hangzhou 310058, China}
\author{Shanyin Fu}
\affiliation{Center for Correlated Matter and School of Physics, Zhejiang University, Hangzhou 310058, China}
\author{Yifu Xu}
\affiliation{Center for Correlated Matter and School of Physics, Zhejiang University, Hangzhou 310058, China}
\author{Pengyue Xiong}
\affiliation{Center for Correlated Matter and School of Physics, Zhejiang University, Hangzhou 310058, China}
\author{Jiawen Zhang}
\affiliation{Center for Correlated Matter and School of Physics, Zhejiang University, Hangzhou 310058, China}
\author{Mengwei Xie}
\affiliation{Center for Correlated Matter and School of Physics, Zhejiang University, Hangzhou 310058, China}
\author{Yanan Zhang}
\affiliation{Center for Correlated Matter and School of Physics, Zhejiang University, Hangzhou 310058, China}
\author{Ye Chen}
\affiliation{Center for Correlated Matter and School of Physics, Zhejiang University, Hangzhou 310058, China}
\author{Huiqiu Yuan}
\affiliation{Center for Correlated Matter and School of Physics, Zhejiang University, Hangzhou 310058, China}
\affiliation{Institute for Advanced Study in Physics, Zhejiang University, Hangzhou 310027, China}
\affiliation{Collaborative Innovation Center of Advanced Microstructures, Nanjing University, Nanjing 210093, China}
\author{Xin Lu}
\affiliation{Center for Correlated Matter and School of Physics, Zhejiang University, Hangzhou 310058, China}
\affiliation{Collaborative Innovation Center of Advanced Microstructures, Nanjing University, Nanjing 210093, China}
\author{Qing-Hu Chen}
\email {qhchen@zju.edu.cn}
\affiliation{Zhejiang Key Laboratory of Micro-Nano Quantum Chips and Quantum Control, School of Physics, Zhejiang University, Hangzhou 310027, China}
\affiliation{Institute for Advanced Study in Physics, Zhejiang University, Hangzhou 310027, China}
\affiliation{Collaborative Innovation Center of Advanced Microstructures, Nanjing University, Nanjing 210093, China}
\author{Yang Liu}
\email {yangliuphys@zju.edu.cn}
\affiliation{Center for Correlated Matter and School of Physics, Zhejiang University, Hangzhou 310058, China}
\affiliation{Institute for Advanced Study in Physics, Zhejiang University, Hangzhou 310027, China}
\affiliation{Collaborative Innovation Center of Advanced Microstructures, Nanjing University, Nanjing 210093, China}

\date{\today}%
\addcontentsline{toc}{chapter}{Abstract}

\begin{abstract}
Combining superconductivity with spintronics provides exciting opportunities to realize low-dissipation quantum devices. Here we report the synthesis, characterization and magnetotransport measurements of the Pb/Bi$_2$Sr$_2$CaCu$_2$O$_{8+\delta}$ (BSCCO) superconducting heterostructures, where an insulating PbO$_{x}$ layer spontaneously forms at the interface. Non-volatile switching between superconducting (logical ``0”) and normal (``1”) states in Pb films by an external field, i.e., infinite magnetoresistance (IMR), can be realized and are attributed to the strong trapping and pinning of vortices in BSCCO. Furthermore, butterfly-shaped hysteresis loops in magnetoresistance, pronounced resistance dips/jumps and thermal reset to superconducting states can be observed and are direct manifestations of the peculiar vortex dynamics in BSCCO and vortex coupling across the Pb/BSCCO interface. Our work demonstrates a simple and effective way to realize IMR through superconducting vortices and opens up new opportunities to study the vortex interactions across the superconducting interfaces. 
\end{abstract}

\maketitle

\section{INTRODUCTION}

The discovery of giant magnetoresistance (GMR) in magnetic layered structures kick-started the field of spintronics \cite{GMR1,GMR2}. The ability to manipulate electron spin and achieve large magnetoresistance enables revolutionary developments and widespread applications in magnetic sensors and non-volatile memory devices, such as hard disks and Magnetoresistive Random Access Memory (MRAM) \cite{gallagherDevelopmentMagneticTunnel2006}. In recent years, there have been considerable interests in combining spintronics with superconductivity \cite{linderSuperconductingSpintronics2015}, where the intrinsic zero-resistance state in superconductors offers an extreme limit of magnetoresistive response \cite{IMRtheory1,IMRtheory2,SFS,komoriMagneticExchangeFields2018}. Indeed, \del{several experimental works have demonstrated} the infinite magnetoresistance (IMR) \add{effect was first experimentally demonstrated} in ferromagnet/superconductor/ferromagnet (FM/SC/FM) heterostructures \cite{IMR1,IMR2,matsukiRealisationGennesAbsolute2025}: the superconducting state of the SC layer can be switched on or off by changing the relative spin orientation of the FM layers via an external field (SC spin valve effect), leading to the pronounced IMR. The capability of binary switching and non-volatile control between superconducting and normal-resistive states provides the basis for superconducting magnetic memory devices, where dissipationless transport of superconductors can be fully utilized. 

In addition to spin valves, Abrikosov vortices, which carry a quantized flux quantum and are ubiquitous in type-II superconductors, can also be used to control the superconductivity and realize the IMR. Such vortex-induced IMR has been demonstrated in Nb-based Josephson junctions, where the stray field or the phase shift caused by the vortex can give rise to fine changes in the Fraunhofer modulation of the critical current, leading to large magnetoresistance \cite{golodSingleAbrikosovVortices2015,golodWordBitLine2023}. Since the vortex can be easily manipulated by magnetic field, electrical current and even light \cite{golodSingleAbrikosovVortices2015,golodWordBitLine2023,Veshchunov2016}, such vortex-based superconducting devices are promising for real applications. Nevertheless, how the vortex-induced response can take place across the interface between two dissimilar superconductors, particularly for high-temperature superconductors, remains to be studied. It is also important to develop a simple and effective way to switch the superconductivity and realize IMR in such heterostructures.  

In this work, we study the vortex-controlled superconductivity and vortex interaction across the Pb/Bi$_2$Sr$_2$CaCu$_2$O$_{8+\delta}$ (BSCCO) interface. Such interface was investigated extensively in the past for phase-sensitive tunneling measurements \cite{wollmanExperimentalDeterminationSuperconducting1993,mossleAxisJosephsonTunneling1999}, although there was much less study on the details of the interface. Instead of studying the Josepheson effect, we here focus on the magnetoresistance measured from the Pb films on top and its hysteretic behaviors. We found that an insulating PbO$_{x}$ layer spontaneously forms at the interface, which facilitates a straightforward investigation of the superconductivity of Pb in response to the vortex of the BSCCO substrate. The peculiar vortex dynamics of BSCCO and their interaction with Pb vortices lead to the nonvolative switching of IMR in Pb films, as well as fine structures in the magnetoresistance. Our work not only opens up new opportunities to utilize the vortex in high-temperature superconductors for superconducting spintronics, but also offers a simple (yet sensitive) transport-based approach for studying vortex dynamics and interactions across interfaces.

\section{EXPERIMENTAL METHODS}

High-quality optimally doped BSCCO single crystals ($T_{c} \sim$ 91 K) were grown by G. D. Gu using the floating-zone method \cite{guLargeSingleCrystal1993}. After cleaving the BSCCO crystals in the ultrahigh vacuum (UHV) chamber, Pb films were grown at room temperature using a home-built molecular beam epitaxy (MBE) system. The base pressure of the MBE chamber is better than $5 \times 10^{-10}$~mBar. The evaporation of Pb was achieved using an effusion cell, whose deposition rate was set to $\sim 3~\text{\AA}$/min, determined by a quartz crystal microbalance (QCM) monitor. The growth was monitored by $in$-$situ$ reflection high-energy electron diffraction (RHEED) measurements, where streaky RHEED patterns can be observed from both the substrate and Pb films. 

The XRD measurements were performed at room temperature in a Rigaku Ultima IV diffractometer with Cu K$\alpha$ photons (with minor contributions from Cu K$\beta$ photons). Magnetization measurements were carried out using a Magnetic Property Measurement System (MPMS, Quantum Design). Transport measurements were performed using the standard four-probe method in a Physical Property Measurement System (PPMS, Quantum Design), where the four electrodes were made of Au wires that were carefully attached onto the Pb film surfaces using silver paste [Fig.\ref{Fig1}(a)]. 

For photoemission measurements, another MBE system that is connected to a photoemission system was employed. The Pb/BSCCO samples were grown under similar conditions and were then transferred under UHV to the photoemission system, which was equipped with a helium lamp for $in$-$situ$ photoemission measurements. The sample temperature was kept at $\sim$ 6~K during photoemission measurements. The base pressure of the ARPES system was $6 \times 10^{-11}$~mBar, which increased
to $2.5 \times 10^{-10}$~mBar after the helium lamp operation. The energy and momentum resolutions are $\sim$ 12 meV and $\sim 0.01~\text{\AA}^{-1}$, respectively.

\section{Results and Discussion}
\subsection{FABRICATION AND CHARACTERIZATION OF THE Pb/BSCCO HETEROSTRUCTURE}

\begin{figure}
  \centering
  \includegraphics[width=1\columnwidth]{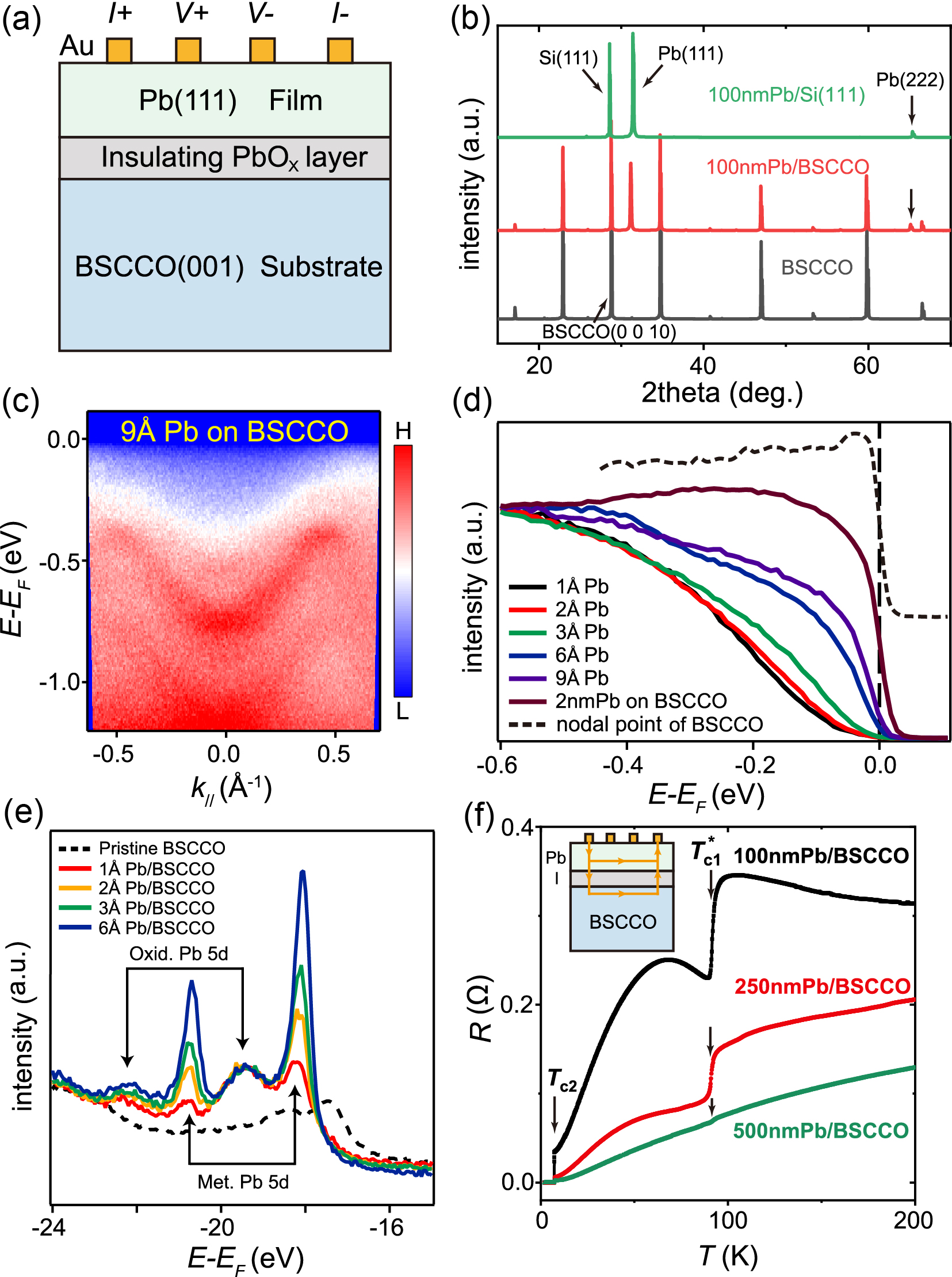}
  \caption{Characterization of the Pb/BSCCO heterostructure, showing the formation of an insulating PbO$_{x}$ layer at the interface.
  	(a) Sideview of the heterostructure. Transport measurements were performed using four Au electrodes on top.
  	(b) XRD scans of the BSCCO substrate (black), the Pb/BSCCO heterostructure (red) and a reference Pb/Si(111) sample (green).
  	(c) ARPES spectra of a 9 {\AA} Pb film ($\sim$3 monolayers) grown on BSCCO, obtained using 21.2 eV photons. 
  	(d) Momentum-integrated energy distribution curves (EDCs) near the Fermi level ($E_F$), obtained at the different stages of Pb deposition.
  	(e) Shallow core-level scans of Pb films with different thicknesses grown on BSCCO. Pb $5d$ peaks from both Pb oxides and metallic Pb are labelled.
  	(f) Temperature-dependent resistance of three Pb/BSCCO heterostructures with different Pb thicknesses. The inset shows the current flow paths (orange arrows).
  	}
  \label{Fig1}
\end{figure}

To fabricate high-quality Pb/BSCCO heterostructures, Pb thin films with different thicknesses were grown epitaxially under ultrahigh vacuum on optimally doped, vacuum cleaved BSCCO ($T_c$ $\sim$ 91K) crystals at room temperature (RT), using molecular beam epitaxy (MBE). (001)-oriented large BSCCO crystals (with a typical in-plane size of 3x3 mm$^2$ \add{and thickness of $\sim$100 $\mu$m}) were chosen for our studies. X-ray diffraction (XRD) measurements [Fig.~\ref{Fig1}(b)] revealed sharp and well-defined Pb(111) and Pb(222) peaks from the Pb films, similar to those of Pb films grown epitaxially on Si(111) substrates. The electron diffraction patterns obtained during the growth further show that the sample surface is flat and smooth [Fig. S1 in supplementary information \cite{SI}]. These results imply that despite the four-fold symmetry of the BSCCO(001) substrate, epitaxial Pb(111) films with a three-fold symmetry can be successfully grown, with two different domains that are rotated $90^\circ$ along $z$ with each other. 

Measurements of the valence bands near the Fermi level ($E_F$) [Fig.\ref{Fig1}(c)], obtained from $in$-$situ$ angle-resolved photoemission spectroscopy (ARPES), clearly revealed the characteristic quantum well states (QWSs) of the ultrathin Pb films due to quantum confinement effects along the $z$ direction\cite{ChiangPhotoemission,QW2}. Direct observation of QWSs indicates that the Pb film is smooth and has a sharp interface with the underlying BSCCO substrate, which is facilitated by the formation of an insulating PbO$_{x}$ layer at the interface. The insulating interfacial layer can be verified by the momentum-integrated energy distribution curves (EDCs) from ARPES measurements shown in Fig.\ref{Fig1}(d): upon deposition of small amount of Pb, e.g., 1-3 {\AA}, the density of states (DOSs) at $E_F$ is largely suppressed and exhibits typical insulating behavior. This is in stark contrast to the quasiparticle peak of pristine BSCCO at the nodal point, as well as the well-defined Fermi edge for thicker Pb films [see Fig.\ref{Fig1}(d)]. The insulating interfacial layer is most likely made of PbO$_{x}$, which can be inferred from the core level scans shown in Fig.\ref{Fig1}(e). Here the thickness-dependent scans reveal that, at the early stage of Pb deposition, the Pb core levels already consist of two sets of spin-orbit split Pb $5d$ peaks (5$d_{3/2}$ and 5$d_{5/2}$): the set with lower energy corresponds to PbO$_{x}$ \cite{PbBSCCOSoftxrayPhotoemission1992,PbYBCOPhotoemission1990}, and its intensity does not increase much upon further deposition below 6 {\AA}; the set with higher energy can be attributed to metallic Pb, whose intensity increases obviously for thicker Pb films. Therefore, based on these $in$-$situ$ photoemission results, we conclude that an insulating PbO$_{x}$ layer spontaneously forms at the interface [see Fig.\ref{Fig1}(a)], due to the oxidation of Pb at the Pb/BSCCO interface. \add{And the PbO$_{x}$ layer is limited to a few interfacial layers and metallic Pb becomes dominant for subsequent Pb deposition. }

The spontaneous formation of an insulating layer at the interface has interesting implications. First, it indicates that direct deposition of a metallic superconductor (here Pb) onto the BSCCO substrate can naturally yield a superconductor/insulator/superconductor (S/I/S') Josepheson junction with a sharp interface. This is a necessary condition for observing the Josepheson effects in the $c$-axis BSCCO junctions \cite{DURUSOY1996253,mossleAxisJosephsonTunneling1999,PhysRevLett.84.2235,PhysRevLett.132.017002}. Second, the formation of the interfacial insulating layer can partly explain why the search for the superconducting proximity effect from BSCCO is challenging in electron spectroscopy \cite{shimamuraUltrathinBismuthFilm2018}, in addition to the well-known issue of the small coherence length of BSCCO. Last but not least, the finite resistance from the insulating interfacial layer allows for detection of two superconducting transitions from both Pb and BSCCO using the standard transport measurement [see Fig.\ref{Fig1}(a)]: here the measured resistance is basically the resistance of the Pb film connected in parallel with the thin interface and the BSCCO substrate \cite{PhysRevLett.102.147002}, and the associated current flow is demonstrated in the inset of Fig.\ref{Fig1}(f). Indeed, the measured resistance $R$ clearly shows two distinct transitions at $T_{c1}^*$ $\sim$ 90 K and $T_{c2}$ $\sim$ 7 K [Fig.\ref{Fig1}(f)], which is very close to the superconducting transition temperatures of bulk BSCCO and Pb thick film [see Fig. S2 in supplementary information \cite{SI}], respectively. With increasing thickness of Pb, the two transitions remain discernible, although the transition at $T_{c1}^*$ appears weaker due to the reduced resistance from Pb films. \del{The finite resistance from the thin insulating interface ensures that the superconducting transition of the Pb films can be detected.} 

\add{We emphasize that the insulating interface prevents the superconducting BSCCO substrate from electrically shorting the circuit of the Pb film, allowing for the observation of the magnetoresistance from the Pb overlayer. The temperature-dependent resistance for the 100 nm Pb film exhibits an upturn below $\sim$ 90 K and a subsequent decrease below $\sim$ 60 K [Fig.\ref{Fig1}(f)]. The resistance upturn can be attributed to defect scattering from the interface. A similar upturn has been reported in cuprates when point defects are introduced by electron irradiation \cite{PhysRevLett.91.047001}. Indeed, the upturn is progressively suppressed with increasing Pb thickness, reflecting the growing contribution from the metallic conducting path from the Pb overlayer. Given the very thin PbO$_{x}$ layer estimated from core level scans [Fig.\ref{Fig1}(e)], interfacial defects are most likely limited to a few atomic layers near the interface. On the other hand, the hysteretic magnetoresistance, to be reported below, takes place at temperatures well below the resistance hump at $\sim$ 60 K and is governed by the vortex dynamics of the underlying BSCCO substrate.}

\subsection{VORTEX-INDUCED INFINITE MAGNETORESISTANCE}

\begin{figure*}
	\centering
	\includegraphics[width=2\columnwidth]{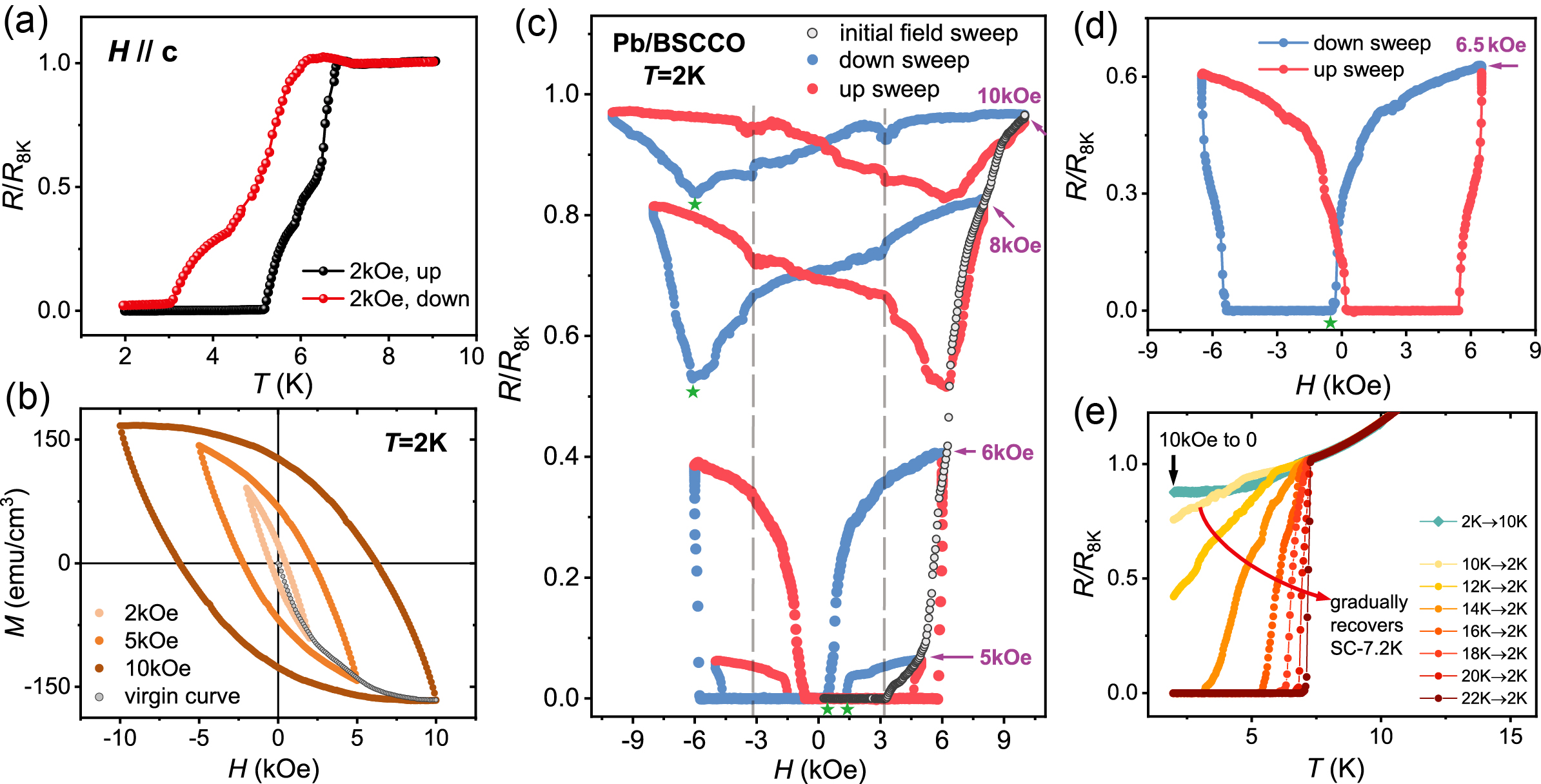}
	\caption{Magnetoresistance and its hysteresis of the Pb/BSCCO heterostructure.
		(a) Normalized resistance of a 100 nm Pb/BSCCO heterostructure as a function of temperature for a warming and cooling cycle under an out-of-plane magnetic field of 2 kOe. A freshly grown sample was first cooled to 2 K without a magnetic field, after which a field of 2 kOe was applied at 2 K before this measurement.
		(b) Magnetization–field ($M$–$H$) hysteresis loops of bulk BSCCO measured at $T$=2 K. Each loop corresponds to a specific maximal applied field (MAF) and its value is indicated on each curve. The grey circles indicate the initial $M$–$H$ curve starting from zero field (virgin curve).
		(c) Resistance–field ($R$-$H$) hysteresis loops at 2 K for different MAFs (their values are labelled in purple). The black circles indicate the $R$-$H$ curve during the initial field sweep, i.e., starting from a fresh superconducting state without trapped vortices. \add{The green stars highlight the field $H_\text{Rmin}$, where the zero resistance (for low and intermediate MAF) or the minimal resistance (for high MAF) is just achieved during the downward field sweep. $H_\text{Rmin}$ is positive for low MAF and negative for intermediate MAF, see Fig.~\ref{Fig4}(a-b).} Vertical dashed lines mark the field $H_\text{c2,eff}$, where the (local) vortices in Pb films undergo dramatic changes [see Fig.~\ref{Fig4}(d) and related discussions]. 
		(d) The $R$-$H$ hysteresis loop under a judiciously chosen field of 6.5 kOe, where non-volatile switch of IMR can be realized. 
		(e) Temperature-dependent resistance under zero external field after sequential warming-cooling cycles to elevated temperatures (as indicated on each curve). The sample was initially non-superconducting at 2 K, due to the application and removal of a 10 kOe field.
	}
	\label{Fig2}
\end{figure*}

\begin{figure*}
	\centering
	\includegraphics[width=2\columnwidth]{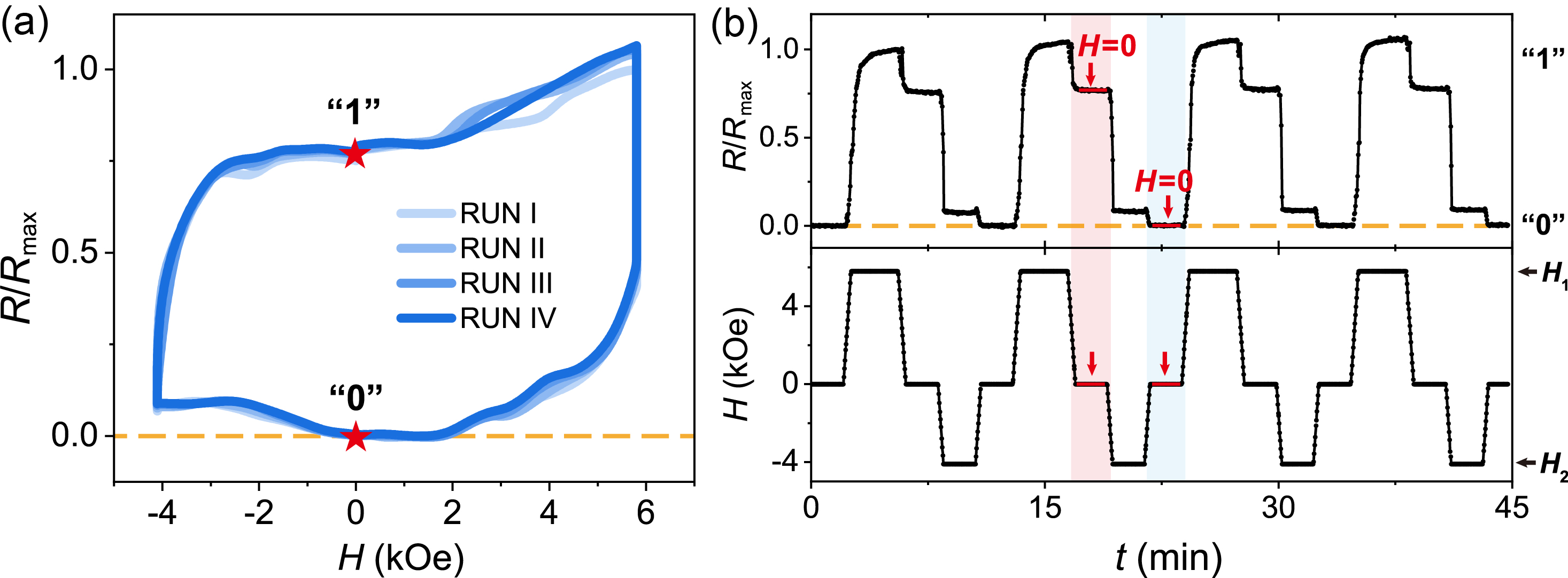}
	\caption{Demonstration of the magnetic switching of superconductivity in Pb and associated IMR. 
		(a) The $R$–$H$ hysteresis loops for a 250 nm Pb/BSCCO heterostructure. The finite-resistance state at zero field corresponds to the ``1" state, while the superconducting state is the ``0" state, as highlighted by red stars.
		(b) Demonstration of repetitive switching between the ``0" and ``1" states by an external field. Bottom: time-dependent magnetic field; top: the corresponding resistance. The light red and blue regions indicate the ``1" and ``0" states, respectively. }
	\label{Fig3}
\end{figure*}

A key finding of our paper is the vortex-induced IMR in the Pb/BSCCO heterostructure, as demonstrated in Fig.~\ref{Fig2} and~\ref{Fig3}. After growth, the sample was first zero-field cooled from RT to 2 K, followed by applying an out-of-plane magnetic field of 0.2 T. A subsequent warming and cooling cycle reveals a clear hysteresis in resistance, as shown in Fig.~\ref{Fig2}(a). Such hysteresis disappears if the BSCCO substrate is replaced by Si, or the field is applied within the $a$-$b$ plane [see Supplementary Informationin \cite{SI}, Fig. S3]. These facts indicate that the BSCCO substrate is critical for the observed hysteresis: applying a magnetic field $H$ along the $c$-axis induces a vortex lattice in the BSCCO substrate ($H>H_{c1}$), which acts as an effective magnetic field for the Pb film. In contrast, for fields applied within the $a$-$b$ plane, the field lines from the BSCCO vortices have negligible impact on the Pb film on top. For this reason, all the data shown below were obtained for fields applied along the $c$-axis. Fig.~\ref{Fig2}(b) displays the $M$–$H$ hysteresis loops of BSCCO measured at $T$= 2~K, where a substantial remanent magnetization remains after removal of the external field, due to the well-known vortex trapping \cite{chenRevealingMicroscopicMechanism2024,minkovMagneticFluxTrapping2023}. 

In our S/I/S' heterostructure, the remanent magnetic field from the BSCCO vortex can control the superconducting state of Pb through a carefully controlled external field, resulting in IMR that can be utilized for superconducting spintronics \cite{IMR1,IMR2,matsukiRealisationGennesAbsolute2025,golodWordBitLine2023}. Fig.~\ref{Fig2}(c) and \ref{Fig2}(d) show the hysteretic magnetoresistance in a typical 100-nm Pb/BSCCO heterostructure. Initially, the sample was in a clean superconducting state after zero-field cooling to 2 K, where no (or very little) vortex trapping takes place. Gradually increasing the magnetic field [along the initial field sweep, i.e., the black circles in Fig.~\ref{Fig2}(c)] leads to suppression of superconductivity in Pb at $\sim$3 kOe and rapid growth of resistance beyond this field. Fig.~\ref{Fig2}(c) shows a few $R$–$H$ hysteresis loops in representative maximal applied fields (MAFs), which are marked by purple arrows with the numerical values indicated. For low MAFs, e.g., 5 and 6 kOe, clear hysteresis loops can already be observed in the $R$–$H$ curves, although superconductivity with zero resistance can be recovered after the field is removed; however, for high MAFs such as 8 and 10 kOe, zero resistance cannot be recovered by simply sweeping the fields, leading to a butterfly-shaped hysteresis loop. Note that all the hysteresis loops are symmetric for up and down sweeps of the same magnetic field, consistent with the vortex dynamics inferred from Fig.~\ref{Fig2}(b). \add{Ignoring the trapped magnetic vortices, the vortex density in BSCCO can be estimated by the applied magnetic field, which is the order of 10$^{10}$ cm$^{-2}$ in our case. However, as we shall discuss in details below, trapped vortices in BSCCO have to be considered for a correct interpretation of the experimental results and hence the size of the hysteresis in the $R$–$H$ curves cannot simply scale with the external magnetic field.}

In Fig.~\ref{Fig2}(d), the $R$–$H$ hysteresis loop at 6.5 kOe is presented, which clearly demonstrates the vortex-induced IMR and its nonvolative control in the Pb/BSCCO heterostructure. Starting from the clean superconducting state at zero field (the ``0" state), application and subsequent removal of a 6.5 kOe field result in a finite resistance at zero field (the ``1" state), leading to IMR. Such control can be better demonstrated in Fig.~\ref{Fig3}, where repeative switchings between the ``0" and ``1" states by an external field are shown for a 250-nm Pb/BSCCO heterostructure. Specifically, application of a +6 kOe field from the initial ``0" state leads to a sharp increase in resistance (the ``1" state), which remains high after removal of the field [Fig.~\ref{Fig3}(a)]. Similarly, application of a -4 kOe field from the initial ``1" state leads to a sharp decrease in resistance, which eventually restores the system to the superconducting ``0” state. \add{We mentioned that the $R$-$H$ hysteresis loops and time-dependent responses are related to the sweep rate of the magnetic field [see Fig. S5 in supplementary information \cite{SI}], due to the dynamics of vortex pinning, trapping and annihilation.} These results demonstrate that two distinct resistance states can be stabilized at zero external field and this type of switching can be repeated in many cycles with high fidelity, as shown in Fig.~\ref{Fig3}(b). 

The field-induced normal-resistance ``1" state can be converted back to the superconducting ``0" state by simply warming up to $\sim$20 K, as shown in Fig.~\ref{Fig2}(e) and Fig. S4 \cite{SI}, where a series of $R$-$T$ curves upon warming to elevated temperatures are displayed. With increasing temperature, the original superconducting state of Pb with $T_c$ = 7.2 K gradually recovers. The characteristic temperature of $\sim$20 K is consistent with the reported vortex melting temperature in BSCCO \cite{20K1,20K2,maTemperatureDependenceVortex2011}, providing further evidence for the vortex-induced IMR (see below for more discussions).

\subsection{VORTEX DYNAMICS AND COUPLING}

\begin{figure*}
	\centering
	\includegraphics[width=2\columnwidth]{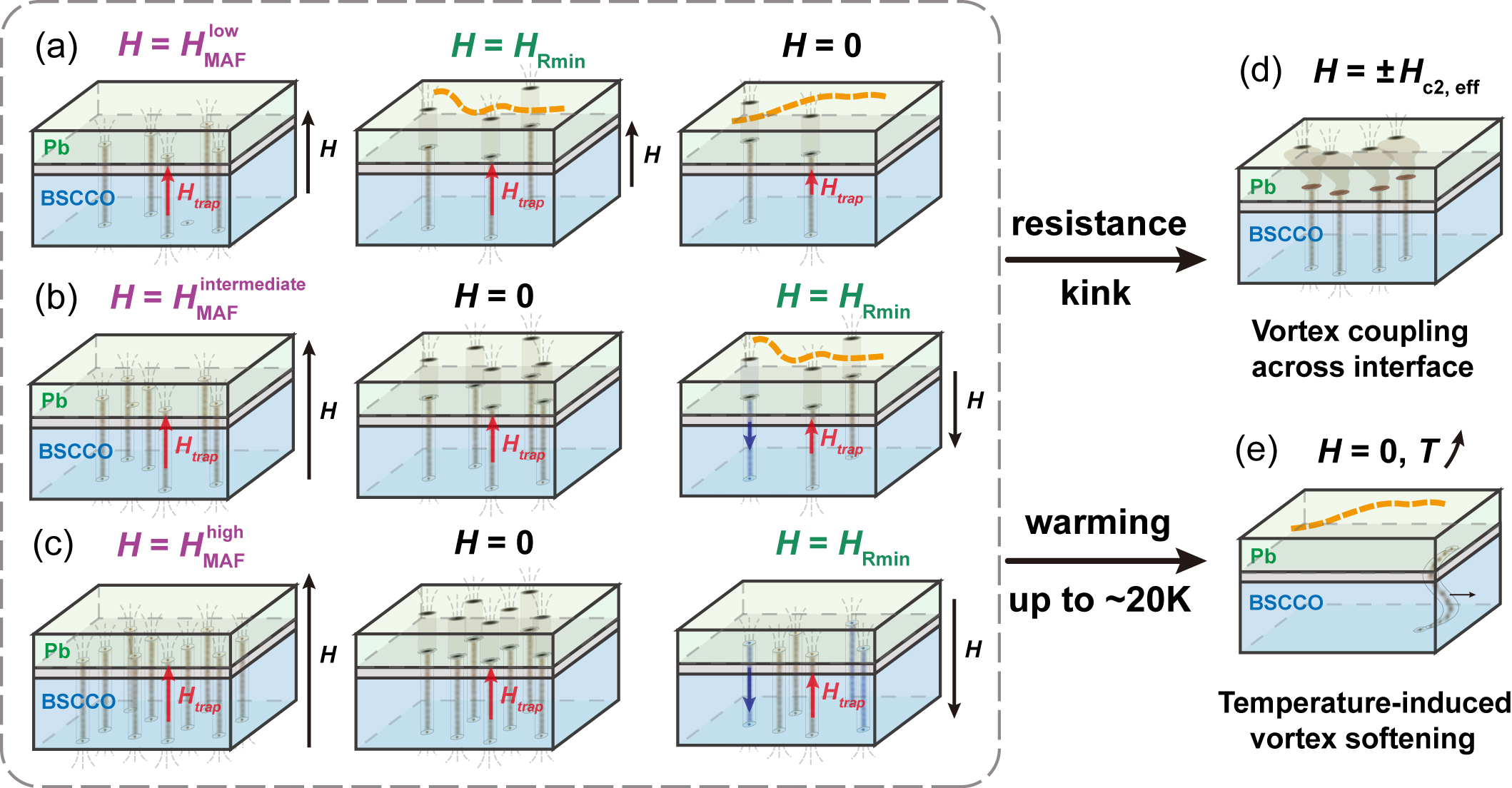}
	\caption{Cartoons illustrating the vortex dynamics that underlies the observed magnetoresistance in the Pb/BSCCO heterostructure. (a-c) Vortices in the Pb/BSCCO heterostructure after an initial field sweep to a low (a), intermediate (b) and high (c) MAF. The three panels (from left to right) indicate vortices at different stages during a downward sweep from the corresponding MAF. $H_\text{Rmin}$ represents the field that has zero or minimal resistance during the downward field sweep [see green stars in Fig.~\ref{Fig2}(c) and \ref{Fig2}(d)]. For low MAF (a), $H_\text{Rmin}$ is positive; for higher MAFs (b-c), $H_\text{Rmin}$ is negative. Red and blue arrows indicate vortices and anti-vortices in BSCCO, respectively. Orange dashed lines in (a,b) indicate the superconducting paths within the Pb films. 
		(d) Vortex coupling between Pb and BSCCO near $H = \pm H_\text{c2,eff}$ [vertical dashed lines in Fig.~\ref{Fig2}(c)], which can give rise to the observed dips/jumps in resistance [Fig.~\ref{Fig2}(c)]. 
		(e) Melting and annihilation of BSCCO vortices and antivortices upon warming above 20 K, which can restablish the superconducting in Pb after cooling. 
	\add{Note that here each vortex (or flux line) in BSCCO actually represents a vertical stack of pancake vortices, which is simplified by a vertical cylinder in this schematic plot. The soft vortex cores are represented by brown/blue circles stacked at the center of each cylinder.}} 
	\label{Fig4}
\end{figure*}

The cartoons in Fig.~\ref{Fig4}(a-c) illustrate the vortex dynamics responsible for the observed hysteretic magnetoresistance shown in Fig.~\ref{Fig2}(c,d). As discussed above, the $R$–$H$ hysteresis loop depends on the MAF ($H_\text{MAF}$) during the initial field sweep and can be classified into three regimes: low MAF $H_\text{MAF}^\text{low}$ [Fig.~\ref{Fig4}(a), corresponding to $<$6 kOe in Fig.~\ref{Fig2}(c)], intermediate MAF $H_\text{MAF}^\text{int}$ [Fig.~\ref{Fig4}(b), corresponding to $\sim$6.5 kOe in Fig.~\ref{Fig2}(d)], and high MAF $H_\text{MAF}^\text{high}$ [Fig.~\ref{Fig4}(c), corresponding to $>$8 kOe in Fig.~\ref{Fig2}(c)]. Since these measurements were performed at low temperature (2 K) and under fields much smaller than $H_{c2}$ of BSCCO, the vortices in BSCCO after the initial field sweep can be regarded as a regular Abrikosov vortex lattice [left panels in Fig.~\ref{Fig4}(a-c)]. As the field is swept downward, the vortex density reduces accordingly. For low MAF $H_\text{MAF}^\text{low}$ [Fig.~\ref{Fig4}(a)], the superconductivity in Pb can be restored at $H_\text{Rmin}$,  before the field reaches zero. Consequently, only a small number of vortices remain trapped inside the BSCCO substrate at zero external field.  Note that Pb films thinner than $\sim$ 1200 nm are found to be type-II superconductors~\cite{dolanCriticalThicknessesSuperconducting1973}, and therefore can also develop vortices. The ``fat" vortices  in Pb (due to its large superconducting coherence length) could be pinned to the ``slim" vortices  in BSCCO (with a short coherence length), allowing the field lines to penetrate along the $z$ direction, as illustrated in Fig.~\ref{Fig4}(a). 

For intermediate MAF \del{$H_\text{MAF}^\text{int}$} \add{$H_\text{MAF}^\text{intermediate}$} [Fig.~\ref{Fig4}(b)], superconductivity cannot be restored after sweeping back to zero field due to trapped vortices. However, zero resistance is recovered by sweeping the field further into the negative direction at $H_\text{Rmin}$ (and beyond). This occurs because applying a negative field introduces antivortices that annihilate the remaining vortices, reducing the overall vortex/antivortex density and thereby reestablishing superconducting paths [rightmost panel in Fig.~\ref{Fig4}(b)].
Therefore, for intermediate MAF, the superconducting (``0") and normal-resistive (``1") states can be precisely controlled by the external field and remain stable after the field is removed, as demonstrated in Fig.~\ref{Fig3}. This paves the way for applications as non-volatile magnetic memories. \add{We mention that IMR can be observed close to $T_{c2} \sim$ 7 K [see Fig. S6 in supplementary information \cite{SI}], where a smaller operational magnetic field is required. This provides an additional control knob to tune the IMR in such system. }

For high MAF $H_\text{MAF}^\text{high}$ [Fig.\ref{Fig4}(c)], the superconductivity of Pb cannot be restored by simply sweeping the field. The resulting $R$–$H$ hysteresis exhibits a characteristic butterfly shape, with the minimum resistance appearing at the negative field $H_\text{Rmin}$ [see data above 8 kOe in Fig.\ref{Fig2}(c)]. The failure to achieve zero resistance here can be attributed to the large number of vortices generated during the initial field sweep, which cannot be effectively removed or annihilated by decreasing or even reversing the external field. At the measurement temperature of 2 K, the vortex creep (or annihilation) rate for BSCCO is suppressed due to strong pinning potentials and insufficient thermal activation \cite{20K1}. As a result, antivortices created by a negative field cannot sufficiently annihilate the existing vortices \cite{grigorievaVortexchainStateInBi2Sr2CaCu2O8+d1995}, leading to a non-superconducting state of Pb where vortices and antivortices coexist. The global minimum resistance at $H_\text{Rmin}$ likely corresponds to a state with the lowest joint density of vortices and antivortices [rightmost panel in Fig.~\ref{Fig4}(c)]. Interestingly, similar butterfly-shaped $R$–$H$ hysteresis has been reported in interfacially bombarded Al/Sr$_2$RuO$_4$ heterostructure \cite{wuDisorderinducedPronouncedMagnetoresistive2025}, which was attributed to the disorder-induced anomalous flux, although the detailed mechanism remains to be confirmed. Here in our Pb/BSCCO case, the well-separated energy/temperature scales between Pb and BSCCO allow us to pin down the vortex as the underlying mechanism for the magnetoresistive hysteresis.

Upon close inspection of the $R$–$H$ hysteresis loops at high fields in Fig.~\ref{Fig2}(c), distinct dips and jumps in resistance can be observed for both downward and upward sweeps at the vertical dashed lines. These correspond to the “effective” upper critical field, $H_\text{c2,eff}$, of the Pb film, where it loses superconductivity during the initial field sweep (black circles in Fig.\ref{Fig2}(c)). Such dips or jumps in resistance can be attributed to the Pb vortex lattice, resembling vortex softening process of peak effect\cite{toft-petersenDecomposingBraggGlass2018,marchevskyTwoCoexistingVortex2001}, which undergoes significant change across its upper critical field ~\cite{bhattacharyaDynamicsDisorderedFlux1993,hendersonMetastabilityGlassyBehavior1996,marleyFluxFlowNoise1995}. Specifically, as the external field decreases and crosses $\pm H_\text{c2,eff}$, some local regions of the Pb film can enter the mixed state, while other areas remain non-superconducting due to the large number of trapped vortices in BSCCO. In the locally mixed region, Pb vortices form and are most likely pinned near the vortex cores of BSCCO \cite{hendersonMetastabilityGlassyBehavior1996,banerjeeAnomalousPeakEffect1998}, as these positions likely minimize the energy cost and preserve magnetic flux continuity across the Pb/BSCCO interface [see Fig.~\ref{Fig4}(d)], leading to a sudden reduction of the resistance. The coupling and decoupling of vortices between Pb and BSCCO therefore provide a natural explanation for the observed dips and jumps in resistance near $H_\text{c2,eff}$, as shown in Fig.~\ref{Fig2}(c).

Fig.~\ref{Fig4}(e) illustrates the vortex dynamics upon warming to $\sim$20 K, which erases the existing vortices and resets the system back to the superconducting ``0" state after cooling [Fig.~\ref{Fig2}(e)]. As temperature increases, trapped vortices can overcome pinning barriers through thermal activation \cite{RzchowskiVortexPinning1990}. This process leads to a progressive melting of the vortex lattice through vortex-antivortex annihilation and the escape of vortices to the sample edge. For BSCCO, $\sim$20 K marks the crossover between the zero-dimensional (individually pinned pancake vortices) and three-dimensional (collectively pinned vortex bundles) pinning regimes ~\cite{20K1,20K2}. In addition, the decay of trapped magnetic vortices in BSCCO follows a logarithmic behavior below $\sim$20 K \cite{maTemperatureDependenceVortex2011}. In our experiments, the Pb/BSCCO system completely expels the trapped vortices after warming above $\sim$20 K, regardless of the initial vortex density [Fig.~\ref{Fig2}(e) and Fig. S4]. However, at temperatures lower than 20 K, the vortex dynamics remain temperature-dependent, causing the $R$–$H$ hysteresis loops in Fig.~\ref{Fig2}(c,d) to vary with measurement temperatures. This provides an additional control knob to manipulate such non-volatile memory devices.

We now compare the vortex-induced IMR here with IMR reported in a superconducting spin valves made of FM/SC/FM heterostructure \cite{IMR1,IMR2,matsukiRealisationGennesAbsolute2025}. While both systems enable non-volatile control of the normal–superconducting transition (\add{and hence similar IMR under an external magnetic field}), the thickness range for reliable operation differs. In FM/SC/FM heterostructure, since the normal-superconducting transition relies on the magnetic exchange interaction between two ferromagnetic layers sandwiching the superconductor, the thickness of the superconducting layer is typically limited to a few nanometers \cite{IMR2,matsukiRealisationGennesAbsolute2025}. By contrast, in the present S/I/S' heterostructure, the normal-superconducting transition depends on vortices whose field lines can extend far along the $z$-axis. This largely relaxes the thickness constraint of the superconducting layer. Indeed, our experiments show that similar IMR can be achieved for films up to 500 nm thick (the thickest Pb films studied in this paper), and likely beyond.

\section{CONCLUSION} 

In summary, we experimentally demonstrate the vortex-induced IMR in the Pb/BSCCO heterostructures, where epitaxial Pb(111) films with the insulating interface with BSCCO spontaneously form. Depending on the magnitude of the MAF, the Pb/BSCCO system exhibits distinct $R$–$H$ hysteresis loops, which can be well understood from the vortex dynamics of BSCCO and the vortex interactions across the interface, as schematically demonstrated in Fig.~\ref{Fig4}. Particularly, in the intermediate MAF, reliable ``0"-``1" transitions and non-volatile memory control can be achieved, paving the way for device applications based on this principle. In addition, the characteristic vortex dynamics in BSCCO provides additional tuning knobs to control the IMR, e.g., resetting to the superconducting ``0" state by a simple temperature cycle. This demonstrates the high tunability and versatility of such vortex-based superconducting device. 

While our current Pb/BSCCO system provides a proof of principle for such vortex-based memory device, further improvements with more suitable materials, e.g, employing top superconducting layer with lower critical field \cite{wuDisorderinducedPronouncedMagnetoresistive2025} and a bottom superconductor with stronger vortex pinning/trapping, could significantly enhance the device properties. In addition, such S/I/S' heterostructure provides a new platform to investigate the vortex interactions across the interface between two superconductors.

\section*{ACKNOWLEDGMENTS}
This work is supported by the National Key R$\&$D Program of China (Grant No. 2023YFA1406303, No. 2022YFA1402200, No. 2024YFA1408900) and the National Science Foundation of China (No. 12525408, 12174331\add{, 92565201}). We thank Guowei Yang, Hao Zheng, Ze Pan, Teng Hua, Xinying Zheng and Prof. Ming Shi for help on experiments and useful discussions. The high-quality BSCCO single crystals were from G. D. Gu. 

\section*{DATA AVAILABILITY}
\del{The supplementary information of this article is available~\cite{SI}.} \add{The data that support the findings of this study are openly available \cite{Zhu_figshare_PbBSCCO}.}

\bibliography{Pb-BSCCO}
\end{document}